\documentclass[aps,prl,reprint,longbibliography,superscriptaddress,nofootinbib,preprintnumbers]{revtex4-2}

\usepackage{amsmath,amssymb}
\usepackage{dsfont}
\usepackage{slashed}
\usepackage{graphicx}
\usepackage{bm}
\usepackage{tikz}
\usepackage[compat=1.1.0]{tikz-feynman}
\usetikzlibrary{decorations.pathmorphing,decorations.markings}
\usepackage[colorlinks=true,allcolors=blue]{hyperref}

\definecolor{lime}{HTML}{A6CE39}
\DeclareRobustCommand{\orcidicon}{\hspace{-1.8mm}
	\begin{tikzpicture}
	\draw[lime, fill=lime] (0,0) 
	circle [radius=0.16] 
	node[white] at (-0.007,-0.007) {{\fontfamily{qag}\selectfont \tiny \,ID}};
	\draw[white, fill=white] (-0.058,0.095) 
	circle [radius=0.005];
	\end{tikzpicture}
	\hspace{-2.5mm}
}

\foreach \x in {A, ..., Z}{\expandafter\xdef\csname orcid\x\endcsname{\noexpand\href{https://orcid.org/\csname orcidauthor\x\endcsname}
			{\noexpand\orcidicon}}
}
\begin{document}

\preprint{MSUHEP-26-011}

\title{Quantum Tomography of Top Quarks as a Probe of\\
Charge-Parity Violation}

\author{Eren Erdogan\orcidA{}}
\affiliation{Department of Physics and Astronomy, Michigan State University,
East Lansing, Michigan 48824, USA}
\affiliation{Department of Physics, Illinois State University, Normal, Illinois 61790, USA}

\author{Kirtimaan A.\ Mohan\orcidB{}}
\affiliation{Department of Physics and Astronomy, Michigan State University,
East Lansing, Michigan 48824, USA}
\author{Keping Xie\orcidC{}}
\affiliation{Department of Physics and Astronomy, Michigan State University,
East Lansing, Michigan 48824, USA}
\affiliation{State Key Laboratory of Dark Matter Physics \& Key Laboratory for Particle Astrophysics and Cosmology (MOE) \& Shanghai Key Laboratory for Particle Physics and Cosmology, Tsung-Dao Lee Institute \& School of Physics and Astronomy, Shanghai Jiao Tong University, Shanghai 201210, China\looseness=-1}
\author{Marcel Yanez\orcidD{}}
\affiliation{Department of Physics and Astronomy, Michigan State University,
East Lansing, Michigan 48824, USA}

\date{\today}

\begin{abstract}
LHC measurements now reconstruct all fifteen parameters of the $t\bar t$
two-qubit spin density matrix, which amounts to a full quantum tomography of the pair. We show that
this data constrains CP violation in the top-Yukawa coupling. The coupling
enters the density matrix at one loop and produces spin correlations that are
odd under CP and do not affect the cross section. Using the first complete
renormalized one-loop density matrix and the experimental covariance, we
obtain complementary constraints comparable to those from direct tree-level $t\bar t H$ and $tH$
production.
\end{abstract}

\maketitle

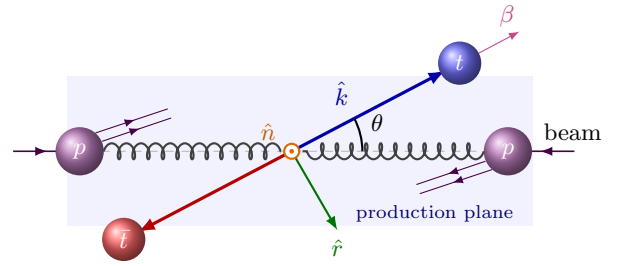
\begin{figure}[t]
  \centering
  \begin{tikzpicture}[>=latex,scale=1.1]
    % production plane and beam axis
    \fill[blue!5] (-2.7,-0.9) -- (2.9,-0.9) -- (2.9,0.9) -- (-2.7,0.9) -- cycle;
    \node[blue!45!black,font=\scriptsize] at (1.72,-0.75) {production plane};
    \draw[gray!55,thin,dashed] (-3.05,0) -- (3.1,0);
    % incoming hadron lines, drawn behind the protons, arrows giving the direction
    \begin{scope}[semithick,violet!55!black,
                  decoration={markings,mark=at position 0.55 with {\arrow{>}}}]
      \draw[postaction={decorate}] (-3.35,0) -- (-2.60,0);
      \draw[postaction={decorate}] (3.40,0) -- (2.65,0);
    \end{scope}
    \node[black,font=\small] at (3.38,0.24) {beam};
    \coordinate (O) at (0,0);
    % incoming protons along the beam
    \shade[ball color=violet!50] (-2.55,0) circle (0.29);
    \node[white,font=\small] at (-2.55,0) {$p$};
    \shade[ball color=violet!50] (2.6,0) circle (0.29);
    \node[white,font=\small] at (2.6,0) {$p$};
    % gluons radiated from the protons, fusing at the interaction point
    \draw[decorate,decoration={coil,aspect=0.55,segment length=5.5pt,amplitude=3pt},
          thick,black!75] (-2.26,0) -- (-0.13,0);
    \draw[decorate,decoration={coil,aspect=0.55,segment length=5.5pt,amplitude=3pt},
          thick,black!75] (2.31,0) -- (0.13,0);
    % proton remnants: parallel spectator lines at a slight angle, mid-line arrows
    % (up-slanted for the left proton, down-slanted for the right)
    \begin{scope}[thin,violet!55!black,
                  decoration={markings,mark=at position 0.55 with {\arrow{>}}}]
      \draw[postaction={decorate}] (-2.28,0.12) -- (-1.45,0.40);
      \draw[postaction={decorate}] (-2.36,0.22) -- (-1.49,0.51);
      \draw[postaction={decorate}] (2.33,-0.12) -- (1.50,-0.40);
      \draw[postaction={decorate}] (2.41,-0.22) -- (1.54,-0.51);
    \end{scope}
    % top and antitop momenta
    \draw[->,very thick,blue!65!black] (O) -- (1.85,0.97);
    \draw[->,very thick,red!70!black] (O) -- (-1.85,-0.97);
    \shade[ball color=blue!55] (2.02,1.06) circle (0.26);
    \node[white,font=\small] at (2.02,1.06) {$t$};
    \shade[ball color=red!60] (-2.02,-1.06) circle (0.26);
    \node[white,font=\small] at (-2.02,-1.06) {$\bar t$};
    % scattering angle and velocity of the top
    \draw[thick,black] (0.85,0) arc (0:27.7:0.85);
    \node[font=\small] at (1.03,0.34) {$\theta$};
    \draw[->,thin,magenta!80!black] (2.25,1.18) -- (2.74,1.44)
      node[above left=-2pt,magenta!80!black,font=\small]{$\beta$};
    % spin-analysis frame {n,r,k}: k lies along the top-quark momentum arrow
    \node[blue!65!black,font=\small] at (0.60,0.70) {$\hat k$};
    \draw[->,thick,green!45!black] (O) -- (0.55,-0.95) node[below,font=\small]{$\hat r$};
    \draw[thick,orange!85!black,fill=white] (O) circle (0.09);
    \fill[orange!85!black] (O) circle (0.028);
    \node[orange!75!black,font=\small] at (-0.28,0.24) {$\hat n$};
  \end{tikzpicture}
  \caption{Kinematics and spin-analysis frame for $t\bar t$ production. Two
    protons ($p$) collide along the beam axis and their gluons, each carrying a
    fraction of the proton momentum, fuse to produce the top (blue) and antitop
    (red), the top emerging at scattering angle $\theta$ with velocity $\beta$ in
    the $t\bar t$ rest frame. The axes are $\hat k$ along the top momentum,
    $\hat r$ in the production plane, and $\hat n$ normal to it.}
  \label{fig:frame}
\end{figure}

Top-quark pairs produced at the LHC have recently become a testing ground for
quantum information at the highest available energies. The spins of the top and
antitop are correlated, and their joint state constitutes a two-qubit system. Since the
top quark decays before it can hadronize, its spin is transferred to its decay
products, whose angular distributions reconstruct the state. Recently,
CMS~\cite{CMS:2024zkc} determined all fifteen independent parameters of the
two-qubit density matrix, a full quantum-state tomography of the produced pair.
This tomographic data has so far been analyzed for the quantum properties of the
state, such as entanglement~\cite{CMS:2024pts}, Bell
nonlocality~\cite{Barr:2024djo}, and magic~\cite{CMS:2025cim,White:2024nuc,Aoude:2025jzc}. We
show here that the same data also constrains Charge-Parity (CP) violation that may be present in the interaction between the top quark
and the Higgs boson.

Such CP violation is well motivated. The excess of matter over antimatter in the
universe cannot be generated with the amount of CP violation present in the
Standard Model (SM), so additional sources are required~\cite{Sakharov:1967dj}. The Higgs boson couples most strongly to the
top quark, so a CP-violating component of
the top-Yukawa coupling is a natural place for such new physics
to appear~\cite{Weinberg:1989dx,Weinberg:1990me,Schmidt:1992et}. We parametrize
this coupling as
\begin{equation}
  \mathcal{L}_{h t\bar t} = -\frac{m_t}{v}\, h\, \bar t\,(a_t + i\, b_t \gamma_5)\, t,
  \label{eq:lagrangian}
\end{equation}
where $a_t$ and $b_t$ are the scalar and pseudoscalar couplings, $v$ is the
Higgs vacuum expectation value, and the SM corresponds to
$(a_t,b_t)=(1,0)$. Any nonzero value of the product $a_t b_t$ violates CP.
Constraining $(a_t,b_t)$ is a central goal of the LHC Higgs
program~\cite{LHCHiggsCrossSectionWorkingGroup:2016ypw}, pursued in
associated $t\bar t H$ and $tH$ production, where the CP structure imprints directly on the production rate, by both ATLAS~\cite{ATLAS:2020ior,ATLAS:2026dao} and
CMS~\cite{CMS:2020cga,CMS:2022dbt}. Electric dipole moments constrain the same
coupling more strongly, but only under assumptions about the couplings of the Higgs
boson to lighter fermions and the absence of cancellations among
contributions~\cite{Brod:2013cka}.

The effect of the phenomenological model presented in Eq.~\eqref{eq:lagrangian} on $t\bar t$ production at hadron colliders has so
far been only partially explored. Refs.~\cite{Maltoni:2024wyh,Martini:2021uey} calculated the fully renormalized CP-even
differential cross-sections of $t\bar t$ production at the LHC, which are insensitive to CP-violating effects.
Refs.~\cite{Schmidt:1992et,Bernreuther:1993df,Bernreuther:1993hq} identified the sensitivity of the $t\bar t$ spin correlations to a CP-violating
Higgs coupling but retained only the ultraviolet-finite parts of the one-loop structure functions. Here we compute the complete
renormalized one-loop Higgs-induced corrections to $gg,q\bar q\to t\bar t$ from the interaction $\mathcal{L}_{h t\bar t}$ of
Eq.~\eqref{eq:lagrangian} and, for the first time, predict their imprint on all fifteen measured Fano coefficients~\cite{Fano:1983zz,CMS:2024zkc}. Comparing these
predictions directly with the data measured by CMS, we obtain the first tomography-based constraints on the CP-even and CP-odd
components of the top-Yukawa interaction.

\noindent\textit{Quantum tomography at the LHC.} Consider a $t\bar t$ pair produced in a definite partonic process, $gg\to t\bar t$
or $q\bar q\to t\bar t$. Its spin state is described by a production spin density
matrix $R$, which can be built from the helicity amplitude $M_{\lambda\bar\lambda}$ for the
process, with $\lambda$ and $\bar\lambda$ the top and antitop helicities, as
$R_{\lambda\lambda',\bar\lambda\bar\lambda'} = M_{\lambda\bar\lambda}\,
M^{*}_{\lambda'\bar\lambda'}$, summed over the helicities and averaged over the
colors of the incoming partons. The trace of $R$ is proportional to the
differential cross section, and
the normalized matrix $\rho = R/\mathrm{Tr}\,R$ is a genuine two-qubit density
operator of unit trace. Any such operator is fixed by fifteen real numbers, which
are conveniently organized as the Fano coefficients $Q_m=\{B_i^{\pm},C_{ij}\}$~\cite{Fano:1983zz,Afik:2020onf},
\begin{equation}
  \rho = \frac14\Big( \mathds{1}\otimes\mathds{1}
  + B_i^{+}\,\sigma_i\otimes\mathds{1}
  + B_i^{-}\,\mathds{1}\otimes\sigma_i
  + C_{ij}\,\sigma_i\otimes\sigma_j \Big),
  \label{eq:fano}
\end{equation}
with summation over the repeated axis indices $i,j$ implied. Here
$B_i^{+}\equiv P_i$ and $B_i^{-}\equiv\bar P_i$ are the polarizations of the
top and antitop, and the nine numbers $C_{ij}$ are the spin-spin correlations; the
index $i$ runs over the three spin axes $\{\hat n,\hat r,\hat k\}$ defined below.
Measuring the complete set is precisely what is meant by quantum-state tomography of
the pair~\cite{Afik:2020onf}. Note that helicity is the projection of
a particle's spin along its direction of motion; since the top and antitop momenta
are antiparallel, the helicity amplitudes, in particular for the antitop, must be
rotated into the common $\{\hat n,\hat r,\hat k\}$ spin basis in which the
coefficients are reported~\cite{Durupt:2025wuk}.

The orthonormal frame $\{\hat n,\hat r,\hat k\}$ used to resolve the coefficients
is attached to each event in the $t\bar t$ rest frame, as shown in
Fig.~\ref{fig:frame}. We adopt the CMS frame and sign conventions for the
measured coefficients, including the $\mathrm{sgn}(\cos\theta)$ folding that
defines the $\hat n$ and $\hat r$ axes~\cite{CMS:2024zkc}.

The same fifteen $Q_m$ also encode the quantum correlations of the pair. Close
to the production threshold the top and antitop emerge in a spin-singlet state
and are maximally entangled, while at large invariant mass they approach a
spin-triplet configuration; the sign and magnitude of the diagonal correlations
track this evolution. It is this fully reconstructed structure, measured across
the kinematic range, that we now turn to a different purpose, namely constraining
the CP-violating part of the top-Yukawa coupling.

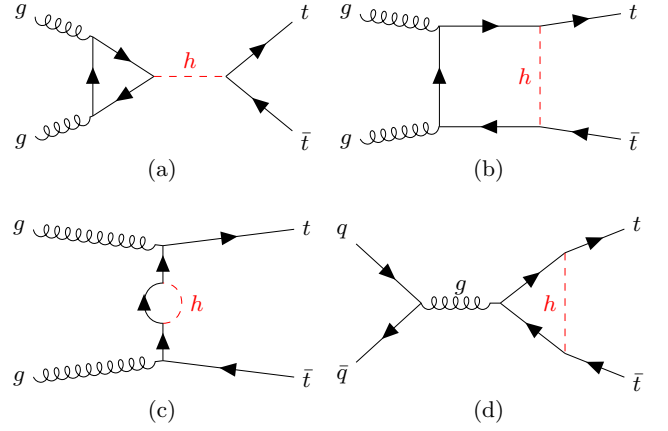
\begin{figure}[t]
  \centering
  \resizebox{\columnwidth}{!}{%
  \begin{tabular}{cc}
    % (a) gg -> h -> ttbar : s-channel Higgs via top triangle
    \begin{tikzpicture}[baseline=(current bounding box.center)]
      \begin{feynman}
        \vertex (g1) at (-2.0, 0.9) {\(g\)};
        \vertex (g2) at (-2.0,-0.9) {\(g\)};
        \vertex (a)  at (-1.0, 0.55);
        \vertex (b)  at (-1.0,-0.55);
        \vertex (c)  at (-0.15, 0.0);
        \vertex (d)  at ( 0.85, 0.0);
        \vertex (t)  at ( 1.95, 0.9) {\(t\)};
        \vertex (tb) at ( 1.95,-0.9) {\(\bar t\)};
        \diagram*{
          (g1) -- [gluon] (a),
          (g2) -- [gluon] (b),
          (a)  -- [fermion] (c) -- [fermion] (b) -- [fermion] (a),
          (c)  -- [scalar, red, edge label=\(h\)] (d),
          (d)  -- [fermion] (t),
          (tb) -- [fermion] (d),
        };
      \end{feynman}
    \end{tikzpicture}
    &
    % (b) gg -> ttbar box with Higgs rung
    \begin{tikzpicture}[baseline=(current bounding box.center)]
      \begin{feynman}
        \vertex (g1) at (-2.0, 0.9) {\(g\)};
        \vertex (g2) at (-2.0,-0.9) {\(g\)};
        \vertex (a) at (-0.7, 0.7);
        \vertex (b) at ( 0.7, 0.7);
        \vertex (c) at ( 0.7,-0.7);
        \vertex (d) at (-0.7,-0.7);
        \vertex (t)  at ( 2.0, 0.9) {\(t\)};
        \vertex (tb) at ( 2.0,-0.9) {\(\bar t\)};
        \diagram*{
          (g1) -- [gluon] (a),
          (g2) -- [gluon] (d),
          (tb) -- [fermion] (c),
          (c) -- [fermion] (d),
          (d) -- [fermion] (a),
          (a) -- [fermion] (b),
          (b) -- [fermion] (t),
          (b) -- [scalar, red, edge label'=\(h\)] (c),
        };
      \end{feynman}
    \end{tikzpicture}
    \\[2pt]
    {\small (a)} & {\small (b)} \\[10pt]
    % (c) top self-energy from a Higgs loop on the exchanged top
    \begin{tikzpicture}[baseline=(current bounding box.center)]
      \begin{feynman}
        \vertex (g1) at (-2.0, 1.05) {\(g\)};
        \vertex (t)  at ( 2.0, 1.05) {\(t\)};
        \vertex (a)  at ( 0.0, 0.80);
        \vertex (b)  at ( 0.0, 0.28);
        \vertex (c)  at ( 0.0,-0.28);
        \vertex (e)  at ( 0.0,-0.80);
        \vertex (g2) at (-2.0,-1.05) {\(g\)};
        \vertex (tb) at ( 2.0,-1.05) {\(\bar t\)};
        \diagram*{
          (g1) -- [gluon] (a),
          (a) -- [fermion] (t),
          (b) -- [fermion] (a),
          (c) -- [fermion, half left, looseness=1.6] (b),
          (c) -- [scalar, red, half right, looseness=1.6, edge label'=\(h\)] (b),
          (e) -- [fermion] (c),
          (g2) -- [gluon] (e),
          (tb) -- [fermion] (e),
        };
      \end{feynman}
    \end{tikzpicture}
    &
    % (d) qqbar -> ttbar with t-channel Higgs exchange (vertex correction)
    \begin{tikzpicture}[baseline=(current bounding box.center)]
      \begin{feynman}
        \vertex (q)  at (-2.0, 1.0) {\(q\)};
        \vertex (qb) at (-2.0,-1.0) {\(\bar q\)};
        \vertex (a)  at (-0.9, 0.0);
        \vertex (b)  at ( 0.2, 0.0);
        \vertex (c)  at ( 1.1, 0.7);
        \vertex (d)  at ( 1.1,-0.7);
        \vertex (t)  at ( 2.1, 1.1) {\(t\)};
        \vertex (tb) at ( 2.1,-1.1) {\(\bar t\)};
        \diagram*{
          (q)  -- [fermion] (a),
          (a)  -- [fermion] (qb),
          (a)  -- [gluon, edge label=\(g\)] (b),
          (b)  -- [fermion] (c),
          (d)  -- [fermion] (b),
          (c)  -- [scalar, red, edge label'=\(h\)] (d),
          (c)  -- [fermion] (t),
          (tb) -- [fermion] (d),
        };
      \end{feynman}
    \end{tikzpicture}
    \\[2pt]
    {\small (c)} & {\small (d)} \\
  \end{tabular}%
  }
  \caption{Representative one-loop Higgs-exchange diagrams for $t\bar t$
    production that generate its dependence on the top-Yukawa couplings
    $(a_t,b_t)$: (a)
    $s$-channel exchange through a top-quark triangle, (b) box with an internal
    Higgs, (c) top self-energy from a Higgs loop, and (d) $q\bar q$ vertex
    correction with $t$-channel Higgs exchange.}
  \label{fig:loopdiagrams}
\end{figure}

\noindent\textit{Quantum imprints of CP violation.}
At leading order in the strong coupling the $t\bar t$ spin state is generated
by pure QCD: the production density matrix is P- and CP-even, and the helicity
amplitudes $M_{\lambda\bar\lambda}$ develop no discontinuities across physical
thresholds. The couplings $(a_t,b_t)$ first enter at one loop, through the
interference of the tree-level QCD amplitude with virtual Higgs exchange;
representative diagrams are shown in Fig.~\ref{fig:loopdiagrams}. Since both
particles in the loop are massive, this correction is infrared finite, and it
induces a definite shift $\Delta R(a_t,b_t)$ of the production density matrix
of relative order $y_t^2/(16\pi^2)$, with $y_t=\sqrt{2}\,m_t/v$ the top-Yukawa
coupling.

The shift $\Delta R(a_t,b_t)$ separates into two physically distinct pieces,
distinguished by whether intermediate states go on shell. The absorptive
part collects all such on-shell contributions, namely the discontinuities of
the amplitude across its branch cuts, and induces imaginary contributions to
the helicity amplitudes.
The two pieces populate different Fano coefficients, in a pattern fixed by
their transformation properties under CP and naive time reversal $T_N$
(reversal of all momenta and spins, without interchange of initial and final
states). The CP-even coefficients, among them the diagonal spin
correlations, receive corrections proportional to $a_t^2$ and $b_t^2$,
whereas every CP-odd coefficient is linear in the product $a_t b_t$ and thus
vanishes unless both couplings are present. The CP-odd set splits further:
the antisymmetric correlations $C^{-}_{nr}$ and $C^{-}_{nk}$ (with
$C^{-}_{ij}=C_{ij}-C_{ji}$) are odd under $T_N$ and can be generated by the
dispersive parts alone, whereas the polarization difference $P_k-\bar P_k$ is
$T_N$-even; being odd under the combined CPT$_N$ transformation, it is forced
by CPT invariance to be proportional to the absorptive
part~\cite{Bernreuther:2015yna}. A nonzero $P_k-\bar P_k$ therefore requires
both CP violation and on-shell intermediate states: it probes CP violation
mediated specifically by particles light enough to be produced on shell, in
this case the top quark and the Higgs boson of Fig.~\ref{fig:loopdiagrams}.
As the results below show, the Fano coefficients most sensitive to
$(a_t,b_t)$ are $P_k$, $\bar P_k$ and $C^{-}_{nr}$.

\noindent\textit{Renormalization of the Spin Density Matrix.}
Predicting the complete set of Fano coefficients requires renormalization of
their dispersive parts, which are ultraviolet divergent; earlier
studies~\cite{Schmidt:1992et,Bernreuther:1993df,Bernreuther:1993hq} retained only the
ultraviolet-finite contributions. We renormalize the one-loop amplitudes in
the on-shell scheme~\cite{Denner:2019vbn}, which requires the top-quark mass
and field counterterms extracted from the top-quark
self-energy~\cite{Denner:2019vbn,Fontes:2021iue}. The CP-even contributions
to the self-energy exhibit the same Lorentz structure as in the SM,
containing vector and scalar components, and have been determined earlier in
Refs.~\cite{Martini:2021uey,Maltoni:2024wyh}. By contrast, the CP-odd
contribution proportional to $a_t b_t$ induces an additional pseudoscalar
structure in the self-energy, which can be renormalized in more than one
way. Here, we use a scheme where the top-quark field renormalization
constants $\delta Z^L_t$ and $\delta Z^R_t$ pick up imaginary components
that are proportional to $a_t b_t$~\cite{Fontes:2021iue}; no new independent
counterterms are required. To our knowledge, this yields the first complete
renormalized production density matrix, with both the absorptive and
dispersive CP-odd observables available for comparison with experiment. We
provide additional details of the renormalization in the Appendix.

\noindent\textit{From amplitudes to a constraint.} We determine the production density matrix directly from the scattering
amplitudes using the kinematic approach of Ref.~\cite{Cheng:2024rxi}. Each
partonic configuration is characterized by the top-quark velocity $\beta$ and
the scattering angle $\theta$ in the $t\bar t$ rest frame
(Fig.~\ref{fig:frame}). The Higgs-induced one-loop contribution enters as an
additive correction to the production density matrix,
\begin{equation}
R^{\alpha}(a_t,b_t)
=
R^{\alpha,\rm SM}
+
\Delta R^{\alpha}(a_t,b_t),
\label{eq:R}
\end{equation}
where
$
\Delta R^{\alpha}(a_t,b_t)
=
R^{\alpha,h t\bar t}(a_t,b_t)
-
R^{\alpha,h t\bar t}(1,0)
$
denotes the shift relative to the SM point
$(a_t,b_t)=(1,0)$ and is computed from the one-loop Higgs-induced amplitudes
shown in Fig.~\ref{fig:loopdiagrams}. Here
$\alpha=\{gg,q\bar q\}$ labels the partonic production channel.

The quantity $R^{\alpha,\rm SM}$ represents the complete SM
prediction: the leading-order QCD contribution of order $\alpha_s^2$ together
with its higher-order QCD, electroweak, and SM Higgs corrections, the last
being the $(a_t,b_t)=(1,0)$ point of Fig.~\ref{fig:loopdiagrams}, of relative
order $y_t^2/(16\pi^2)$. Rather than recomputing these contributions, we take
the corresponding Fano coefficients from the SM prediction reported by CMS,
computed at next-to-leading order (NLO) in QCD matched to parton showers and
normalized to the next-to-next-to-leading-order (NNLO)
cross section, with first-order electroweak corrections accounted for in the
quoted uncertainties~\cite{CMS:2024zkc,HEPData:153301}, and add the
Higgs-induced shift $\Delta R^\alpha(a_t,b_t)$. Further, since CMS presents unfolded results, we compare with
them directly, without parton showering, hadronization, or decay simulation; the
production density matrix used here and the one reconstructed from the decay angular
distributions coincide for quantities linear in the density matrix, such as the Fano
coefficients, while nonlinear functions of it, such as entanglement
measures, can differ~\cite{Cheng:2024rxi}.

CMS reports Fano coefficients in bins of $m_{t\bar t}$ and $\cos{\theta}$. For each bin $k$ we integrate the $R$ matrix over
$\beta$ and $\cos\theta$ with the parton-luminosity and phase-space weight~\cite{Afik:2020onf}
\begin{equation}
  W^{\alpha}(\beta,\cos\theta) = \frac{dL^{\alpha}}{d\tau}\,
  \frac{\beta^{2}}{16\pi s\,(1-\beta^{2})},
  \label{eq:weight}
\end{equation}
where $dL^{\alpha}/d\tau$ is the parton luminosity at $\tau=m_{t\bar t}^{2}/s$ with
$\sqrt s=13~\mathrm{TeV}$ and the second factor combines the two-body phase space
$\beta/(32\pi m_{t\bar t}^{2})$ with the Jacobian $d\tau/d\beta$. This yields the
integrated production matrix for channel $\alpha$,
\begin{equation}
  \bm{\mathcal{R}}_k^{\alpha}(a_t,b_t)=\int_k W^\alpha\, R^{\alpha}(a_t,b_t)\,d\beta\,d\cos\theta.
  \label{eq:Rk}
\end{equation}
Since the Fano coefficients are defined from the normalized density matrix
(cf.\ Eq.~\eqref{eq:fano}), the partonic channels cannot be combined
coefficient by coefficient. Instead the gluon- and quark-initiated production
matrices, each already weighted by its parton luminosity through $W^\alpha$,
are summed into the total $\bm{\mathcal{R}}_k=\sum_\alpha \bm{\mathcal{R}}_k^{\alpha}$
before the fifteen Fano coefficients are extracted. Gluon fusion dominates: the
quark-initiated contribution to the Higgs-induced correction is suppressed by
roughly a factor of sixteen around the $t\bar{t}$ threshold region, reflecting the smaller quark-antiquark luminosity for $t\bar{t}$ production at the LHC. The total $\bm{\mathcal{R}}_k$ is related to the cross section in the $k$-th bin by
$\mathrm{Tr}[\bm{\mathcal{R}}_k(a_t,b_t)]/4 =\sigma_k= \sigma_k^{\rm SM} + \mathrm{Tr}[\Delta \bm{\mathcal{R}}_k(a_t,b_t)]/4$. Normalizing $\bm{\mathcal{R}}_k$ by its trace
gives the predicted density matrix
\begin{equation}
  \rho_k(a_t,b_t) = \frac{\rho_k^{\rm SM}
  + \frac{\Delta \bm{\mathcal{R}}_k(a_t,b_t)}{4 \sigma_k^{\rm SM}}}
  {1 + \frac{\mathrm{Tr}[\Delta \bm{\mathcal{R}}_k(a_t,b_t)]}{4 \sigma_k^{\rm SM}}}.
  \label{eq:rhopred}
\end{equation}
Here $\rho_k^{\rm SM}$ is the SM prediction taken directly from the
CMS HEPData record~\cite{HEPData:153301}, and
$\sigma_k^{\rm SM}=f_k\times 832^{+40}_{-46}~\mathrm{pb}$, where $f_k$ is the fractional cross
section in the bin and $832~\mathrm{pb}$ is the total $t\bar t$ cross section
computed at NNLO with
next-to-next-to-leading-logarithmic resummation~\cite{Czakon:2013goa}. The
renormalization and factorization scales are set to $H_T/2$, with $H_T$ the
sum of the transverse masses of the top and antitop, the top-quark
mass to $172.5~\mathrm{GeV}$, and parton luminosities are computed from the NNPDF3.1 NNLO parton
distributions~\cite{NNPDF:2017mvq} following the CMS analysis~\cite{CMS:2024zkc}. 
We compute the renormalized one-loop amplitudes and the density matrices built
from them analytically with FeynArts and
FormCalc~\cite{Hahn:2000kx,Hahn:1998yk}, evaluating the loop integrals with
LoopTools~\cite{Hahn:1998yk}. We cross-check them against a fully independent
numerical implementation, in which the renormalized model, including the CP-odd
counterterms, is generated with FeynRules and
NLOCT~\cite{Alloul:2013bka,Degrande:2014vpa} and the virtual amplitudes are
computed with MadGraph5\_aMC@NLO~\cite{Alwall:2014hca,Hirschi:2011pa}. The two
implementations yield identical constraints.

We then extract the
fifteen predicted Fano coefficients $Q_m(a_t,b_t)$ from $\rho_k$ through the
decomposition of Eq.~\eqref{eq:fano}, form the residuals
$\Delta Q_m = Q_m(a_t,b_t) - Q_m^{\rm obs}$, and compare with the measurement
through
\begin{equation}
  \chi^2(a_t,b_t) = \sum_{m,n} \Delta Q_m\,(V^{-1})_{mn}\,\Delta Q_n,
  \label{eq:chi2}
\end{equation}
with $Q_m^{\rm obs}$ the measured Fano coefficients and $V$ their full experimental
covariance matrix, both reported by CMS~\cite{HEPData:153301}.
From the $\chi^2$ of Eq.~\eqref{eq:chi2} we construct three fits,
distinguished by the observables entering it. The first,
$\chi^2_{m_{t\bar t}}$, uses kinematic information alone. The couplings
$(a_t,b_t)$ also shift the $t\bar t$ cross section and its kinematic
distributions~\cite{Maltoni:2024wyh,Martini:2021uey,CMS:2019art}, and the
trace of $\Delta\bm{\mathcal{R}}_k$ gives the Higgs-induced shift of the
differential $m_{t\bar t}$ cross-section. We evaluate this shift in each bin
of the CMS measurement of Ref.~\cite{CMS:2021vhb} and compare with the
measured spectrum and its published covariance through a $\chi^2$ of the
same form as Eq.~\eqref{eq:chi2}. The shift depends on the couplings only
through $a_t^2$ and $b_t^2$ and not on the product $a_t b_t$, so
$\chi^2_{m_{t\bar t}}$ does not probe CP violation. The second,
$\chi^2_{\text{Fano-inc}}$, uses spin information alone, namely the fifteen
Fano coefficients measured inclusively over the full kinematic range, with
their full covariance~\cite{CMS:2024zkc,HEPData:153301}. The third,
$\chi^2_{\text{Fano-diff}}$, combines the spin and kinematic information
within a single measurement. CMS also reports the fifteen Fano coefficients
in four $m_{t\bar t}$ bins, together with a normalization $c$ for every bin
that serves as a proxy for the shape of the differential $m_{t\bar t}$
cross-section, and a covariance correlating all sixty-four
quantities~\cite{CMS:2024zkc,HEPData:153301}. Details of
$\chi^2_{m_{t\bar t}}$ and $\chi^2_{\text{Fano-diff}}$ are given in the
Appendix.

\begin{figure}[t]
  \centering
  \includegraphics[width=\columnwidth]{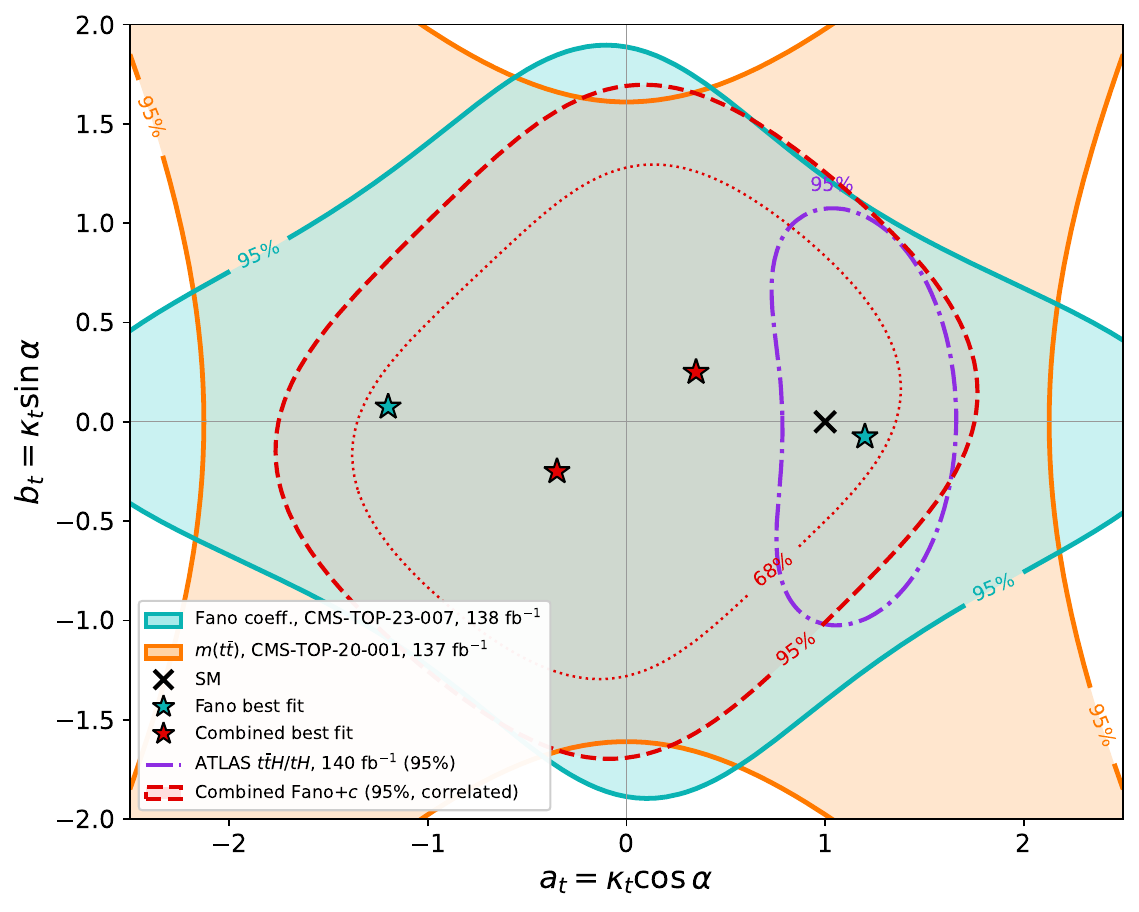}
  \caption{Complementary constraints on the top-Yukawa couplings
    $(a_t,b_t)=(\kappa_t\cos\alpha,\kappa_t\sin\alpha)$ from CMS $t\bar t$ Run~2 data.
    The blue shaded region is the $95\%$ CL allowed by
    $\chi^2_{\text{Fano-inc}}$, the fit to the spin density matrix measured
    in Ref.~\cite{CMS:2024zkc}; the orange shaded region is the $95\%$ CL
    allowed by $\chi^2_{m_{t\bar t}}$, the fit to the differential
    $m_{t\bar t}$ cross-section measured in Ref.~\cite{CMS:2021vhb}. The red
    dashed and dotted contours are the $95\%$ and $68\%$ CL of
    $\chi^2_{\text{Fano-diff}}$, the combined fit of the Fano coefficients
    and the normalizations $c$ for each bin, with correlations from the
    published covariance. The SM is indicated
    by ($\times$) and the stars indicate the degenerate minima of each fit
    involving the Fano coefficients. The ATLAS $t\bar t H/tH$ $95\%$ contour
    from the same Run~2 dataset~\cite{ATLAS:2026dao} is overlaid
    (dash-dotted).}
  \label{fig:contours}
\end{figure}

\noindent\textit{Results.} Figure~\ref{fig:contours} presents the constraints in the $(a_t,b_t)$ plane
from the three fits. Since the one-loop density matrix depends on the
couplings only through $a_t^2$, $b_t^2$, and the product $a_t b_t$, each fit
is invariant under the sign change $(a_t,b_t)\to(-a_t,-b_t)$, which produces
two mirror minima, indicated by stars in Fig.~\ref{fig:contours} for the two
fits involving the Fano coefficients. In all three fits the SM point
$(a_t,b_t)=(1,0)$ is consistent with the data within the 68\% CL limit.

The orange shaded region is the 95\% confidence level (CL) region of
$\chi^2_{m_{t\bar t}}$, which uses only the differential $m_{t\bar t}$
cross-section of Ref.~\cite{CMS:2021vhb}. As expected of a purely kinematic
fit, it constrains the overall magnitude of the couplings but is insensitive
along the $a_t^2\simeq b_t^2$ directions, leaving open the region where $a_t$
and $b_t$ are simultaneously nonzero, which is precisely where CP is
violated.

In contrast, the blue shaded region, the 95\% CL region of
$\chi^2_{\text{Fano-inc}}$, uses only the spin information carried by the
fifteen inclusive Fano coefficients~\cite{CMS:2024zkc} and closes the
$a_t^2\simeq b_t^2$ directions. The best-fit point is
$(a_t,b_t)=(1.20,-0.08)$, and at $a_t=1$ the allowed interval is
$-1.40<b_t<1.23$ at $95\%$ CL. While all Fano coefficients vary in the $(a_t,b_t)$ plane, experimental uncertainties, which determine the sensitivity of each,  are small only for a small number of coefficients. For a reference
CP-violating coupling, $(a_t,b_t)=(1,1)$, we quantify the effect on each
coefficient by the pull $\delta Q_m/\sigma_m$, where $\delta Q_m=Q_m(a_t,b_t)-Q_m^{\rm SM}$
is its predicted shift and $\sigma_m=\sqrt{V_{mm}}$ the uncorrelated per-coefficient uncertainty. The antisymmetric correlation $C_{nr}^{-}$ gives the
largest pull, $0.8\sigma$, followed by the polarizations along $\hat k$, $P_k$
and $\bar P_k$, near $0.6\sigma$, and the diagonal correlation $C_{nn}$ near
$0.4\sigma$;
Fig.~\ref{fig:fano2d} in the Appendix maps this pull for every coefficient
across the $(a_t,b_t)$ plane. 
% The sensitivity concentrates in $P_k$, $\bar P_k$, and
% $C_{nr}^{-}$ because a coefficient constrains the couplings through the ratio
% of its predicted shift to its uncertainty: the polarizations are the most
% precisely measured of the fifteen coefficients, and $C_{nr}^{-}$ is the best
% measured of the antisymmetric correlations. 
The antisymmetric correlation
$C_{nr}^{-}$ carries the dispersive contribution to CP violation, and the
shifts of $P_k$ and $\bar P_k$ carry the absorptive one. The fit therefore
bounds both contributions simultaneously, which cross-section measurements do not
provide.

The fits $\chi^2_{\text{Fano-inc}}$ and $\chi^2_{m_{t\bar t}}$ are therefore
most sensitive along independent directions in the $(a_t,b_t)$ plane, and
combining the spin and kinematic information strengthens the constraint. The
third fit, $\chi^2_{\text{Fano-diff}}$, achieves this combination within a
single measurement; its 68\% and 95\% CL regions are shown as the red dotted
and dashed contours in Fig.~\ref{fig:contours}, with the mirror minima again
indicated by stars. The best-fit point is $(a_t,b_t)=(0.35,0.25)$, and at
$a_t=1$ the allowed interval is $-1.01<b_t<1.26$ at $95\%$ CL
($-0.50<b_t<0.83$ at $68\%$ CL).

We compare this bound with direct probes of the same
coupling. The most stringent direct limits come from associated production of a
Higgs boson with top quarks, in the combination of the  $t\bar t H$ and $tH$ channels, which
constrain the CP-mixing angle of the top-Yukawa
coupling~\cite{ATLAS:2026dao,CMS:2022dbt}. When the comparison is made
two-dimensionally, in the same $(a_t,b_t)=(\kappa_t\cos\alpha,\kappa_t\sin\alpha)$
plane and at the same confidence level, the measurement of
Ref.~\cite{ATLAS:2026dao} from the same Run~2 dataset reaches
$|b_t|\approx1.1$ at $a_t=1$, comparable to the bound obtained here and shown as
the dash-dotted curve in Fig.~\ref{fig:contours}. The tomographic
constraints are independent of and complementary to these direct probes,
and the two can be combined to yield stronger constraints on new
physics in the top-Yukawa interaction.

The three fits presented here serve as a demonstration of the method. The
binning of the tomographic measurement was chosen to map the quantum
properties of the $t\bar t$ state, not to search for the loop-induced
shifts computed here. The same data should therefore yield stronger
constraints with a binning tuned for this purpose, such as smaller
$m_{t\bar t}$ bins near the production threshold, where the Higgs-induced
corrections are largest.

\noindent\textit{Discussion.} We have shown that quantum tomography of top-quark pairs provides a new probe of
the CP structure of the top-Yukawa interaction. Using a complete renormalized
one-loop calculation of the Higgs-induced corrections to the $t\bar t$
production density matrix, we predicted the resulting shifts in all fifteen
Fano coefficients and compared them directly with the density matrix
measured by CMS. Both elements are new: the fully renormalized production
density matrix had not been obtained before, and the tomographic data had not
previously been used to constrain CP violation. This approach uses the quantum state itself as the observable,
complementing conventional cross sections and kinematic distributions.

The sensitivity of tomographic observables is fully realized only after
including the complete Higgs-induced contribution to the production density
matrix, with renormalized dispersive terms. Using the measured Fano
coefficients and their full covariance matrix, we find, for an SM-like
scalar coupling $a_t=1$, $-1.40<b_t<1.23$ at 95\% confidence level, and
$-1.01<b_t<1.26$ from the combined fit $\chi^2_{\text{Fano-diff}}$, a
sensitivity comparable to existing direct probes of Higgs-top
interactions.

These bounds rest on an implicit assumption: any other new physics is heavy
enough to decouple from $t\bar t$ production, so that
Eq.~\eqref{eq:lagrangian} is the only nonstandard interaction entering the
loop; additional new physics in the loop would modify the bounds. Conversely,
since the Fano coefficients are measured in bins of $m_{t\bar t}$ and
$\cos\theta$, the shape of a loop correction across the spectrum is resolved,
so states not far above the $t\bar t$ threshold would leave a distinguishable
imprint; we leave this direction to future work.

The procedure developed here is general and not limited to CP
violation: any new physics that modifies the production
density matrix, through loops or effective
operators~\cite{Aoude:2022imd,Maltoni:2024tul,Aoude:2025jzc}, can be confronted with the
measured tomography and its covariance in the same way, without regenerating
events or repeating the decay simulation. That computationally expensive
step is already contained in the CMS analysis and does not need to be redone for
every model. Furthermore, these results demonstrate that quantum-state tomography can serve not
only as a probe of quantum correlations at high energies, but also as a
precision observable for testing fundamental interactions of the SM and
its possible extensions.

\begin{acknowledgments}
\noindent\textit{Acknowledgments.} We wish to thank Daniel Concha, Dorival Gonçalves, Fabio Maltoni, Scarlett Rebolledo and Cesar Riquelme for useful discussions.  
The work of K.M., K.X.\ and M.Y.\ was supported in part by the National Science
Foundation under Grant No.~PHY-2310497. M.Y.\ gratefully acknowledges financial support for this publication by the Fulbright U.S. Student Program, which is sponsored by the U.S. Department of State and the Chilean Fulbright Commission.
\textit{AI use statement.}---The authors used Claude (Anthropic) to assist with making the numerical analysis faster, with developing plotting codes, and with editing the manuscript. The authors verified AI-generated
code against independent implementations and take full responsibility for the results and the text.
\end{acknowledgments}

\bibliography{refs}

% ============================================================
%  APPENDIX
% ============================================================
\clearpage
\section*{Appendix}

\noindent\textit{Renormalization of the CP-odd sector.} The one-loop Higgs-exchange
correction to the $t\bar t$ amplitude is ultraviolet divergent, and we render it
finite by renormalizing the top-quark mass and field in the on-shell scheme,
following Refs.~\cite{Denner:2019vbn,Fontes:2021iue,Maltoni:2024wyh}. The
counterterms follow from the one-loop top self-energy $\Sigma_t(p)$, the
one-particle-irreducible $t\bar t$ two-point function, which decomposes into a
vector, a scalar, and a pseudoscalar form factor,
\begin{equation}
  \Sigma_t(p) = \slashed{p}\,\Sigma_V(p^2)
  + m_t\,\Sigma_S(p^2) + i\,m_t\,\gamma_5\,\Sigma_P(p^2).
  \label{eq:selfdecomp}
\end{equation}
For the Higgs-exchange loop these read~\cite{Maltoni:2024wyh}
\begin{equation}
  \begin{aligned}
    \Sigma_V &= -\frac{a_t^2+b_t^2}{16\pi^2}\frac{m_t^2}{v^2}\,B_1,\\
    \Sigma_S &= \frac{a_t^2-b_t^2}{16\pi^2}\frac{m_t^2}{v^2}\,B_0,\\
    \Sigma_P &= \frac{a_t b_t}{8\pi^2}\frac{m_t^2}{v^2}\,B_0,
  \end{aligned}
  \label{eq:formfactors}
\end{equation}
with $B_{0,1}\equiv B_{0,1}(p^2;m_t,m_H)$ the standard two-point functions. The
vector and scalar parts are CP even and reproduce, rescaled by the coupling
strengths, the SM form factors~\cite{Martini:2021uey}; the CP-odd pseudoscalar
$\Sigma_P\propto a_t b_t$ appears only when both couplings are present.

Imposing the on-shell conditions, that the renormalized propagator have its
pole at $p^2=m_t^2$ with unit residue, yields a top-quark mass counterterm
that is real and CP even,
\begin{equation}
  \delta m_t = \frac{m_t^2}{16\pi^2 v^2}\Big[
    a_t^2\big(B_0-B_1\big) - b_t^2\big(B_0+B_1\big)\Big]_{p^2=m_t^2},
  \label{eq:massct}
\end{equation}
depending on the couplings only through $a_t^2$ and $b_t^2$. The pseudoscalar
form factor is instead absorbed by the field renormalization: the left- and
right-handed constants acquire imaginary parts proportional to $a_t b_t$,
\begin{equation}
  \mathrm{Im}\,\delta Z_t^{L,R} = \mp\,\widetilde{\mathrm{Re}}\,\Sigma_P(m_t^2)
  = \mp\,\frac{a_t b_t}{8\pi^2}\,\frac{m_t^2}{v^2}\,
    \Big[\widetilde{\mathrm{Re}}\,B_0\Big]_{p^2=m_t^2},
  \label{eq:dZP}
\end{equation}
nonzero only when both couplings are present. Here $\widetilde{\mathrm{Re}}$ keeps the
imaginary parts of the couplings while discarding the absorptive parts of the loop
integrals. An imaginary axial field renormalization is a chiral rotation of the
top field, equivalent to a pseudoscalar $\bar t\,i\gamma_5 t$ mass term: the
CP-violating Yukawa coupling forces it, as required for the consistent
renormalization of a CP-violating two-point function~\cite{Fontes:2021iue}.
This counterterm cancels the ultraviolet divergence of the pseudoscalar
sector; the finite $a_t b_t$ dependence that remains does not affect the
$t\bar t$ cross section~\cite{Maltoni:2024wyh} but enters the spin density
matrix, where the antisymmetric correlations resolve it. The top self-energy
is the only counterterm carrying an $a_t b_t$ dependence, all others being CP
even; the on-shell subtraction implicit in the ultraviolet-finite treatment of
Ref.~\cite{Bernreuther:1993hq} is identical to this counterterm, and we verify
numerically that the two give the same CP-odd density matrix.

\noindent\textit{Constraints from differential $m_{t\bar t}$ cross-sections ($\chi^2_{m_{t\bar t}}$).} The
$m_{t\bar t}$ contour in Fig.~\ref{fig:contours} uses the trace of the same
renormalized one-loop Higgs-induced shift of the production density matrix, which
gives the corresponding shift of the $t\bar t$ cross section. This shift depends on
the couplings only through $a_t^2$ and $b_t^2$, so it is CP even; the calculation
follows Ref.~\cite{Maltoni:2024wyh}, to which we refer for details. We evaluate it in
each bin of the CMS $m_{t\bar t}$ measurement~\cite{CMS:2021vhb}, add it to the
NNLO Standard Model prediction at its leading-order
normalization, and form a $\chi^2$ against the measured spectrum with its
published covariance to obtain the $95\%$ contour.

\noindent\textit{Combined fit ($\chi^2_{\textrm{Fano-diff}}$).} Along with the fifteen Fano coefficients in four $m_{t\bar t}$ bins, CMS also reports a
normalization $c$ for every bin, defined such that its expected value is
unity. The published covariance includes all correlations among the
sixty-four quantities (four $m_{t\bar t}$ bins, fifteen Fano coefficients, and $c$). We predict the normalization in each $m_{t\bar{t}}$ bin as
\begin{equation}
  c(a_t,b_t) = \nu\,\left(1 + \frac{\Delta\sigma_k(a_t,b_t)}{\sigma_k^{\rm SM}}\right)\ ,
\end{equation}
with $\Delta\sigma_k(a_t,b_t) = \sigma_k(a_t,b_t) - \sigma_k(1,0)$
the cross-section shift of the bin, $\sigma_k^{\rm SM}$ as defined below Eq.~\eqref{eq:rhopred}, and $\nu$ a nuisance parameter that absorbs the
overall normalization of the prediction and is allowed to vary in the fit, so that the fit depends on the shape of the $m_{t\bar{t}}$ distribution and not on the differential cross-section, which carries large scale uncertainties.
The coefficients are predicted as before using Eq.~\eqref{eq:rhopred} and the kinematic information is added to the existing $\chi^2$ defined in Eq.~\eqref{eq:chi2} that contains the spin information.
Additional tests varying the
SM normalization have minimal impact on the contour.

\begin{figure*}[t]
  \centering
  \includegraphics[width=\textwidth]{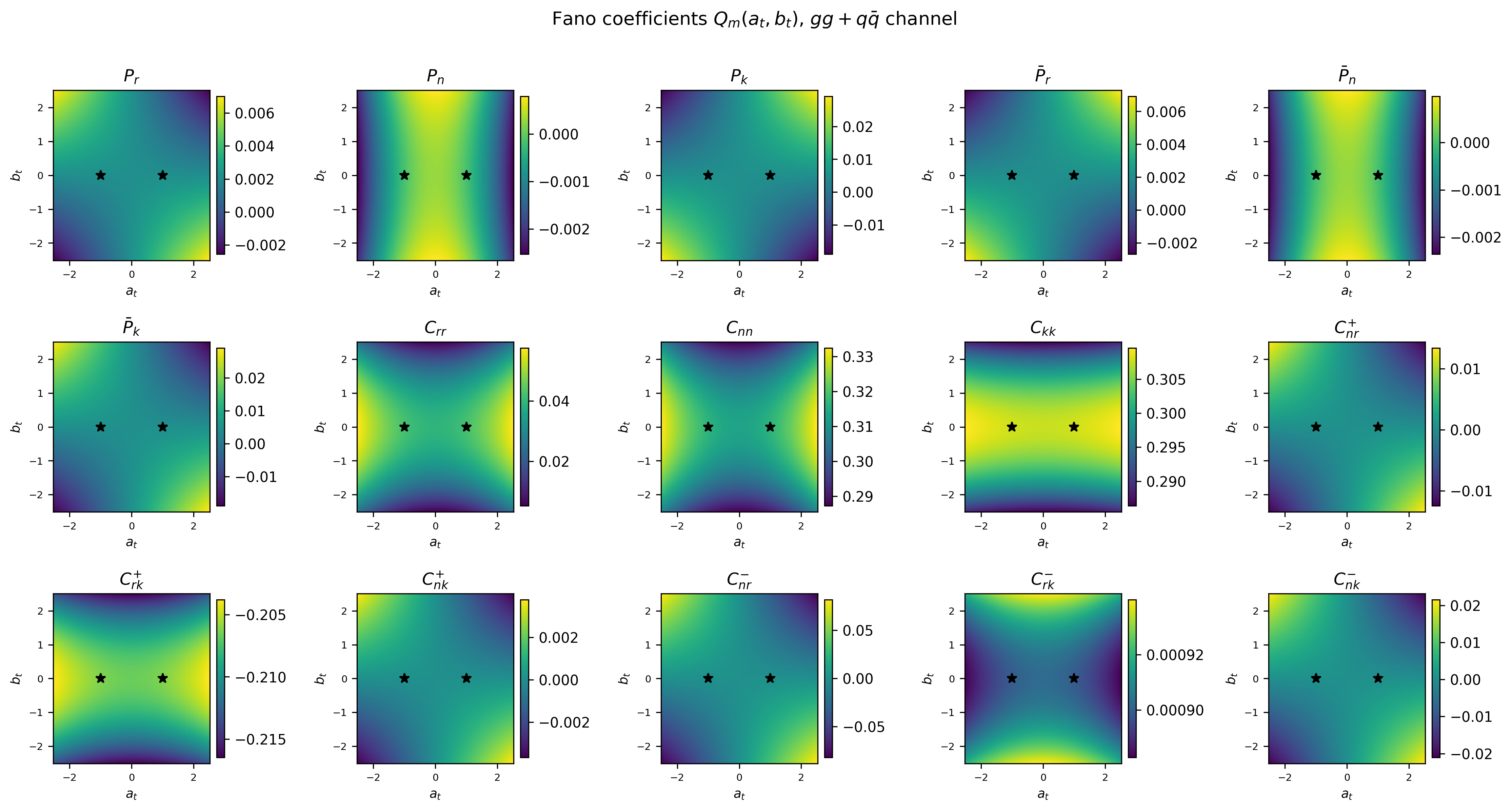}
  \caption{The fifteen Fano coefficients $Q_m(a_t,b_t)$ of the combined
    $gg+q\bar q$ production density matrix across the $(a_t,b_t)$ plane, from the
    complete renormalized one-loop calculation. The stars mark the SM
    point and its mirror.}
  \label{fig:fano2dvals}
\end{figure*}

\begin{figure*}[t]
  \centering
  \includegraphics[width=\textwidth]{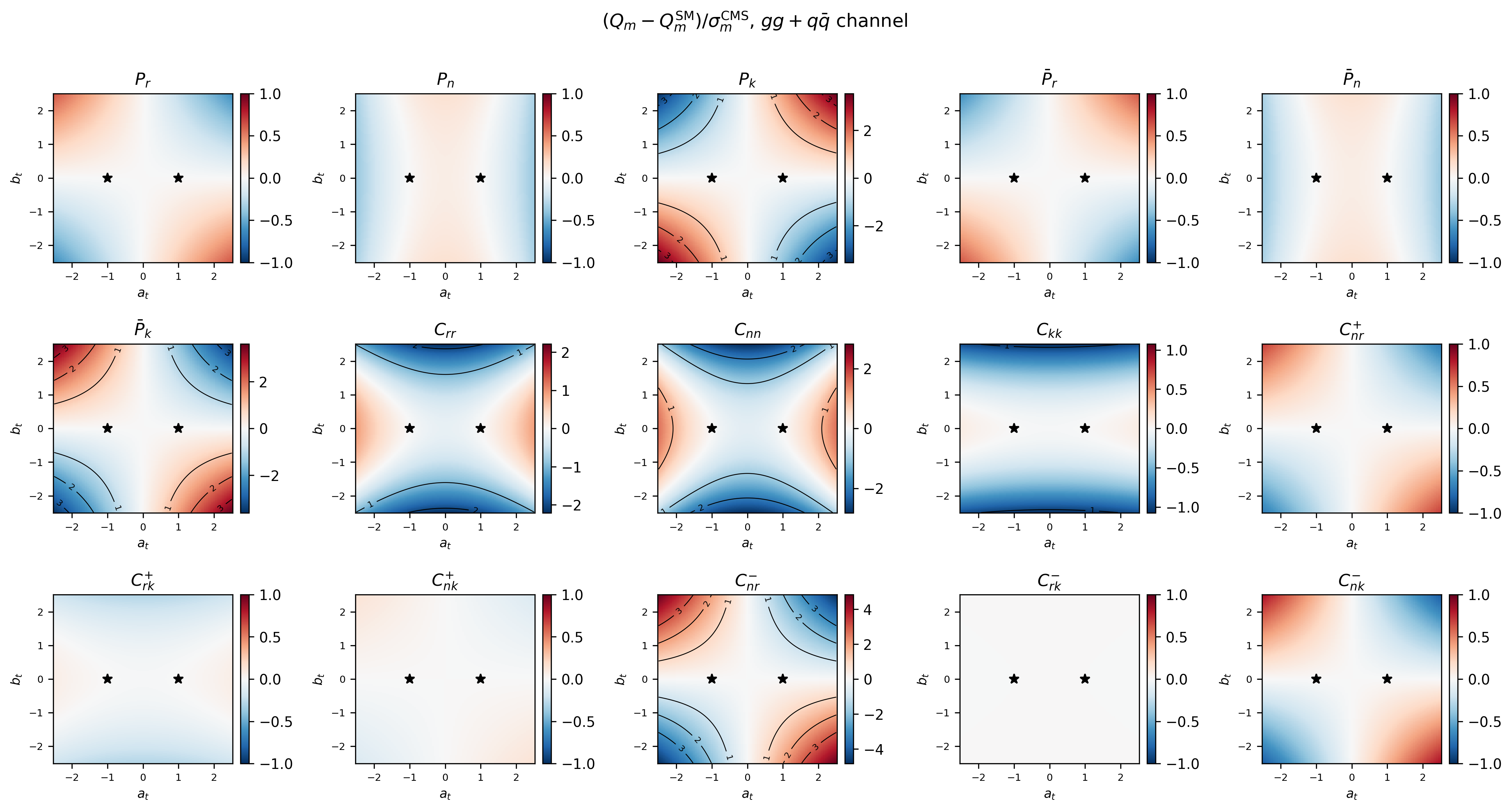}
  \caption{Significance maps of the individual Fano coefficients for the combined
    $gg+q\bar q$ channel: for each of the fifteen Fano coefficients, the predicted
    shift relative to the uncorrelated uncertainty,
    $\delta Q_m/\sigma_m = (Q_m-Q_m^{\rm SM})/\sigma_m$ with $\sigma_m=\sqrt{V_{mm}}$, across the $(a_t,b_t)$
    plane, from the complete renormalized one-loop calculation. Contours mark
    one, two, and three standard deviations; the stars indicate the SM point
    and its mirror. The antisymmetric correlation $C_{nr}^{-}$ shows
    the strongest CP-odd response.}
  \label{fig:fano2d}
\end{figure*}

\noindent\textit{Sensitivity of individual Fano coefficients.}
Fig.~\ref{fig:fano2dvals} shows
the fifteen predicted Fano coefficients $Q_m(a_t,b_t)$ across the plane, and
Fig.~\ref{fig:fano2d} the corresponding pull $\delta Q_m/\sigma_m$ relative to the
uncorrelated uncertainty $\sigma_m$, with contours at one, two, and three standard deviations. The
antisymmetric correlation $C_{nr}^{-}$ carries the strongest CP-odd sensitivity,
consistent with its role in setting the bound in Fig.~\ref{fig:contours}.

\end{document}